\documentclass[twocolumn,pre,floatfix,showpacs]{revtex4}
 
\usepackage{epsfig} 
\usepackage{amsmath} 
\usepackage{times} 
\usepackage[latin1]{inputenc} 
 
\usepackage{ifthen}

\def\be{\begin{equation}} 
\def\ee{\end{equation}}

\begin{document}

\title{Connectivity of Turing structures} 
 
\author{Teemu Lepp\"anen$^{1}$, 
        Mikko Karttunen$^{1}$, 
        R.A. Barrio$^{1,2}$, and 
        Kimmo Kaski$^{1}$} 
 
\address{ $^1$Laboratory of Computational Engineering,  
Helsinki University of Technology,  
P.O. Box 9203, FIN--02015 HUT, Finland \\  
$^2$Instituto de Fisica, 
Universidad Nacional Autónoma de México
(UNAM), Apartado Postal 20-364 01000 
M\'exico, D.F., M\'exico} 
 
\date{\today} 
 
\begin{abstract} 
It is well-known that in two dimensions Turing systems produce spots,  
stripes and labyrinthine patterns, and in three dimensions lamellar  
and spherical structures or their combinations are observed. We study  
transitions between these states in both two and three dimensions by  
first analytically deriving a control parameter and a scaling function  
for the number of clusters. Then, we apply large scale computer simulations 
to study the effect of nonlinearities on clustering, the appearance  
of topological defects and morphological changes in Turing structures.  
In the two-dimensional real space spotty structures we find some evidence 
of twin domain formation, of the kind seen in crystalline materials. With 
the help of reciprocal space analysis we find indication of other more 
general forms of order accommodation, i.e., eutactic local structures. 
Also a mechanism for the observed  ``connectivity transition'' is proposed. 
\pacs{PACS numbers: 82.40.Ck, 47.54.+r, 05.45.-a} 
 
\end{abstract} 
 
\maketitle 
 
\section{Introduction} 
 
Nature presents a fascinating diversity of patterns in plants,  
animals and other natural formations arising often from complex  
physico-chemical processes~\cite{ball}. Alan Turing was the first to  
show that a simple system of coupled reaction-diffusion equations  
for two chemicals could give rise to spatial patterns due to a  
mechanism called diffusion-driven instability~\cite{turing}.  
These so-called Turing patterns have since been proposed to  
account for pattern formation in many biological systems,  
e.g. patterns on fish skin~\cite{kondo,barrio1}, butterfly  
wings~\cite{sekimura} and lady beetles~\cite{liaw}  
to mention a few. However, the first experimental evidence  
of a Turing structure was reported by Castets {\it et al.}~\cite{castets},  
who observed a sustained standing nonequilibrium  
chemical pattern in a single-phase open reactor with  
chloride-iodide-malonic acid (CIMA) reaction. Currently, there  
is an increasing interest to develop simple and plausible  
mathematical models that could describe, at least  
qualitatively, these pattern  
formations~\cite{nicolis,haken,meinhardt,murray}. Although there  
is a variety of models that could potentially produce  
similar patterns, a Turing system is perhaps the simplest one.  
 
The forms and variations of patterns generated by Turing systems  
have been studied by investigating the conditions for 
instability~\cite{szili93a}, assuming inhomogeneous diffusion  
coefficients~\cite{barrio2}, and by introducing domain curvature~\cite{varea1} 
and growth~\cite{varea2}. In addition, symmetries in Turing systems  
are of great interest, since they might have biological relevance, 
see e.g. Refs.~\cite{barrio2,Hasslacher:1993a,aragon}.  
Recently, we have studied the effect of dimensionality by  
simulating three-dimensional Turing systems, which displayed  
complex pattern formation~\cite{teemu}. While in two dimensions  
one obtains spots, stripes or labyrinthine patterns, in three  
dimensions complex shapes of, e.g. lamellae, spherical droplets  
and their combinations appear. 
 
Previously, studies of Turing patterns have typically concentrated  
on reaction kinetic and stability aspects (see 
e.g.~\cite{judd,lengyel:1992a}), while the issue of pattern structure 
and its connectivity has received less attention. Here, we focus 
on connectivity of Turing patterns and its dependence on the parameters 
of the system. We present a simple way to quantitatively characterize 
Turing structures and their connectivity. These methods are not 
only able to characterize the structures, but also to explain some 
of the effects of nonlinearities behind Turing patterns. We also 
demonstrate the existence of a  ``connectivity transition'' when 
the nonlinear interactions of the Turing model are varied.  
 
It was reported earlier that in generic 2D (3D) Turing models  
nonlinear cubic interactions favor stripes (lamellae)  
with both morphogens being connected through the system~\cite{barrio1,teemu}.  
On the other hand, quadratic interactions favor spots (spherical structures),  
in which case only one of the morphogens is connected. Thus, a connectivity  
transition appears when the nonlinear interactions are varied from  
being cubic to predominantly quadratic or vice versa. In order to  
characterize this transition between separated and connected  
structures, we define a dimensionless control parameter and a  
scaling function for the number of clusters. This is tested for a  
variety of system sizes and unstable modes. To characterize the  
resulting morphologies, we calculated the spatial Fourier spectrum to  
get the reciprocal space representation of the chemical concentrations  
and followed its evolution through the connectivity transition.    
 
This paper is organized as follows. Next, we briefly describe  
the reaction-diffusion model of the Turing kind. Then, we discuss 
the concept of connectivity in these systems and the methods for 
characterizing the transition between different patterns. 
In Sec.~\ref{sec:results}, we present results of comprehensive 
numerical simulations, which is followed by Sec.~\ref{sec:kvector} 
with a reciprocal space analysis. Then in Sec.~\ref{sec:concl} we 
draw conclusions.

\section{The Model} 
 
A Turing system models the evolution of the concentrations of 
two chemicals, or morphogens, and is in general 
represented  by the following reaction-diffusion equations 
\begin{eqnarray} 
U_t & = & D_U \nabla^2 U + f(U,V)\nonumber \\ 
V_t & = & D_V \nabla^2 V + g(U,V), 
\label{eq:turing} 
\end{eqnarray} 
where $U \equiv U(\vec{x},t)$ and  $V \equiv V(\vec{x},t)$ are  
the morphogen concentrations, and $D_U$ and $D_V$ the corresponding 
diffusion coefficients setting the time scales for diffusion.  
The reaction kinetics is described by the two nonlinear functions  
$f$ and $g$. 
 
In this study we focus on the generic Turing model 
introduced by Barrio {\it et al.}~\cite{barrio1}, in which 
the reaction kinetics was developed by Taylor expanding  
the nonlinear functions around a stationary solution $(U_c,V_c)$.  
If terms above the third order are neglected, the system reads  
as follows 
\begin{eqnarray} 
u_t & = & D \delta \nabla^2 u + \alpha u(1-r_1 v^2)  + v(1-r_2 u) \nonumber \\ 
v_t & = & \delta \nabla^2 v + v(\beta + \alpha r_1 uv) + u(\gamma + r_2 v), 
\label{eq:barrio} 
\end{eqnarray} 
where $u=U-U_{c}$ and $v=V-V_{c}$ are the concentration fields.  
The parameters $r_1$ and $r_2$ set the amplitudes of the nonlinear  
cubic and quadratic terms, respectively, $D$ is the ratio of the  
diffusion coefficients of the two chemicals, and $\delta$ acts as  
a scaling factor fixing the size of the system. Setting $D \neq 1$  
is a necessary but not a sufficient condition for the diffusion-driven  
instability to occur. For details about the instability and the linear  
stability analysis of the model we refer the reader to Barrio  
{\it et al.}~\cite{barrio1, teemu}. 
 
From the linear stability analysis~\cite{barrio1, teemu} we obtain the  
dispersion relation and the conditions for the diffusion-driven  
instability as the region in $k$-space with positive growth rate, i.e.,  
eigenmodes $u = u_0 e^{\lambda t}$ and $v=v_0 e^{\lambda t}$ with  
eigenvalues $\lambda (k)> 0$. In addition, one can analytically derive  
the modulus of the critical wave vector 
\begin{equation} 
k_c^2 = \frac{1}{\delta} \sqrt{\frac{\alpha (\beta +1)}{D}}, 
\end{equation} 
which was here determined for the case $\alpha=-\gamma$ (set to  
obtain only one stable state at $u=v=0$ in the absence of diffusion).  
In a discretized three-dimensional cubic system, the wave number  
is of the form 
\begin{equation} 
|\vec{k}|= \frac{2 \pi}{L} \sqrt{n_x^2 + n_y^2 + n_z^2}, 
\label{eq:wavevector} 
\end{equation} 
where $L$ is the system size and $n_x$, $n_y$, $n_z$ are the wave  
number indices (in a two-dimensional system $n_z=0$).  
By adjusting the parameters and allowing only a few unstable modes,  
one can obtain several different parameter sets. As in our earlier  
work~\cite{teemu} we chose the parameters $D=0.516$,  
$\alpha =-\gamma =0.899$, $\beta =-0.91$ and $\delta = 2$  
corresponding to a critical wave vector $k_c=0.45$, and $D=0.122$,  
$\alpha =0.398$, $\beta = -0.4$ and $\delta = 2$ corresponding  
to $k_c=0.84$. 
 
In this study, we vary the parameters $r_1$ and $r_2$ in 
Eq.~(\ref{eq:barrio}), since they control the appearance of stripes 
or spots. By  gradually changing these parameters we observe a transition  
from spotty (2D) or spherical droplet (3D) patterns to striped (2D) or  
lamellar (3D) patterns. In order to investigate this transition,  
numerical simulations were carried out by discretizing the spatial  
dimensions into a square or cubic cell lattice and calculating the 
Laplacians in Eq.~(\ref{eq:barrio}). In all of our simulations we used 
$dx = dy = dz = 1.0$. The equations of motion were iterated in time  
using the Euler scheme with time step $dt = 0.05$. The boundary conditions  
were chosen to be periodic, and initially both chemicals were distributed  
randomly over the whole system. 
  
\section{Connectivity} 
 
In the numerical simulations of Eq.~(\ref{eq:barrio}) one deals with 
two concentration fields with characteristic wave lengths. In order to 
visualize this, the concentration of only one of the chemicals is 
typically plotted with a gray scale, since in this type of a system 
the fields are in anti-phase, i.e., if there is a large amount of 
chemical $U$ in some sub-domain, the concentration of chemical $V$ 
would be low there. These concentration fields vary continuously having 
diffuse boundaries. What do we mean by connectivity in patterns of 
chemicals? 
 
To answer this question one can define sub-domains dominated  
by either chemical $U$ or $V$, provided that the amplitudes of  
the patterns is large enough. If we define the boundary as the  
interface between sub-domains dominated by different chemicals,  
we can easily locate the boundaries, since the concentrations  
change rapidly, typically within one or two lattice sites.  
Now, if two points belong to the same domain, i.e., are not separated  
by a boundary, they are considered connected. The definition of  
the boundaries in this way is conceptual in the sense that in the  
$U$-dominated domains the concentration of $V$ does not have to  
be zero, only much less than the concentration of $U$. 
 
\begin{figure} 
\epsfxsize=.9\columnwidth \epsfbox{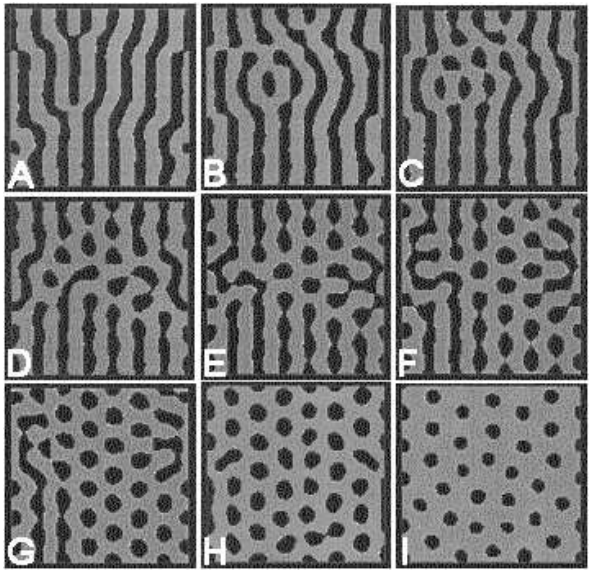}  
\caption{Transition from stripes to spots. The patterns obtained 
after $50000$ iterations in a $100\times100$ system with  
$k_c = 0.45$. Black corresponds to areas dominated by chemical  
$U$ (zeros) and the lighter color chemical $V$ (ones).
Note that the difference in parameters between the figures is 
not constant: From A to I, 
$P=22000$, $120$, $75$, $65$, $60$, $55$, $35$, $15$, $1$.} 
\label{fig:sweep} 
\end{figure}

In Fig.~\ref{fig:sweep} we show changes in the concentration 
fields, i.e., $u$ and $v$, of a 2D system for different values of  
nonlinear parameters $r_1$ and $r_2$ in Eq.~(\ref{eq:barrio}).  
When the cubic term ($r_1$) dominates, the resulting stationary  
pattern is striped with a small number of imperfections, see  
Fig.~\ref{fig:sweep}A. These imperfections can be considered  
as topological defects, or dislocations, which could serve as  
nucleation sites for spots. More dislocations appear  
(see Figs~\ref{fig:sweep}B-C) when the strength of the quadratic term  
is made larger. As the quadratic term increases, more spots  
nucleate and they arrange themselves to triangular structure and  
at the same time getting rid of the remaining stripes  
(see Figs.~\ref{fig:sweep}F-H). Finally when the cubic term is  
diminished even further, only spots remain. In this sequence  
of simulations the strength of the cubic term was changed  
relative to the quadratic term by using a single control  
parameter $P$, which we elaborate below. Nevertheless, 
the transition from striped to spotty pattern seems to happen 
quite abruptly in $P$. In the last frame of this sequence 
(Fig.~\ref{fig:sweep}I) we see a fully stabilized spotty pattern 
with almost perfect triangular symmetry in two different orientations 
such that there is a mirror plane or {\it twin boundary} between 
them.
   
Now, let us discuss patterns in Fig.~\ref{fig:sweep} from the 
clustering point of view. In order to simplify this, and without 
loss of generality, we can assign zeros and ones to the whole  
lattice based on the chemical which dominates a given domain.  
With this mapping we can consider the number of clusters, which can 
be calculated using the well-known Hoshen-Kopelman algorithm~\cite{hoshen} 
as in typical percolation problems. However, before applying this 
cluster algorithm let us first visually inspect possible shapes of 
extended clusters (Fig.~\ref{fig:sweep}).  In Fig.~\ref{fig:sweep}A one 
can see that in the case of stripes the number of $U$- and $V$-dominated 
clusters is almost the same, and both types are extended dominantly in 
one of the dimensions. On the other hand, in the case of a spotty 
structure (Fig.~\ref{fig:sweep}I), chemical $U$ appears as separate 
round clusters or spots, whereas chemical $V$ forms only one connected 
cluster. Between these two limiting cases there is the transition 
region, depicted in Figs.~\ref{fig:sweep}D-F, where $U$-dominated  
clusters appear as spots and stripes in the form of a ``string-of-pearls''. 
A question arises whether the transition from stripes to spots could 
be characterized in a reasonable way by the number of clusters? With the 
Hoshen-Kopelman algorithm we can directly determine the number of clusters ($N$) 
in the system,  but we also need to define a single control parameter 
that $N$ is a function of.  
 
One should bear in mind that many different parameter pairs of 
nonlinear terms ($r_1$ and $r_2$) result in similar patterns. To obtain 
some general insight into the dynamics, we apply dimensional 
analysis~\cite{barenblatt} and derive a dimensionless control parameter 
for describing this behavior. The relevant dimensions are as follows 
\begin{equation} 
[\alpha] = \frac{1}{s},\quad [r_1] = \frac{1}{[c]^2},\quad [r_2] = \frac{1}{[c]s}, 
\end{equation} 
where $s$ denotes seconds and $[c]$ is an arbitrary unit of  
concentration. Thus, the control parameter $P$ can be written in  
the following dimensionless form 
\begin{equation} 
P = r_1 ( \frac{\alpha}{r_2})^2. 
\label{eq:P} 
\end{equation} 
The two limits of this parameter, i.e., $P \to \infty$ ($r_2 \to 0$)  
and $P \to 0$ ($r_1 \to 0$) yield stripes and spots  
(Figs \ref{fig:sweep}A and ~\ref{fig:sweep}I), respectively, as 
the two extremes in 2D. The same holds for 3D. Between these two limits  
a certain combination of striped (lamellar) and spotty (spherical  
droplet) patterns is expected to coexist (see Figs~\ref{fig:sweep}D- 
\ref{fig:sweep}F). 
 
\section{Simulation Results} 
\label{sec:results} 
 
\begin{figure} 
\includegraphics[width=.8\columnwidth, angle=270]{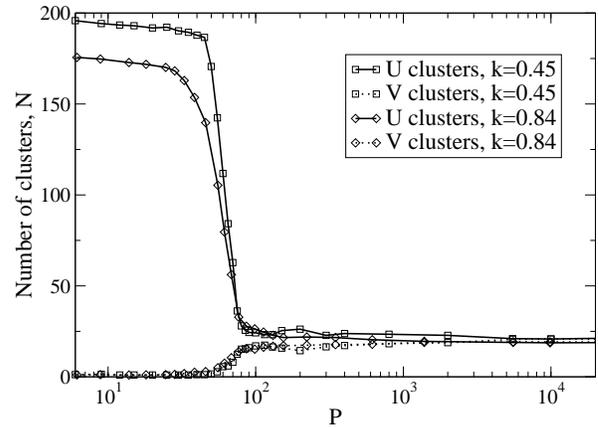} 
\caption{The number of $U$ (solid line) and $V$ (dotted line)  
clusters as a function of the control parameter $P$. 
Diamonds: $k_c=0.84$ in a $100\times100$ system, 
squares: $k_c=0.45$ in a $200\times200$ system. In both cases  
the results were averaged over 20 simulations.} 
\label{fig:result} 
\end{figure}

In order to study the connectivity in two- and three-dimensional  
Turing patterns, we have performed extensive simulations using 
system sizes up to $5\times10^5$ lattice cells and  
up to $2\times10^6$ time steps to reach a stationary state. 
For all cases the results are taken as statistical averages of 
at least 20 separate runs. Using $P$ (Eq.~(\ref{eq:P})) as a control 
parameter is plausible, since the transition from a striped (lamellar) 
pattern to a spotty (spherical droplet) pattern took place in a 
very narrow region of $P$ regardless of the individual values of the 
nonlinear parameters ($r_1$ and $r_2$), the unstable mode 
(dictated by $\alpha$ and $\beta$) or the system size. This is clearly 
seen in Fig.~\ref{fig:result}, where the number of $U$- and $V$-dominated 
clusters is shown as a function of the control parameter $P$ for 
two different cases in 2D. In our studies the $P$-space was 
scanned by keeping either $r_1$ or $r_2$ constant and by  
carrying out separate simulations for each value of $P$. 
From  Fig.~\ref{fig:result} it is clearly seen that the mode 
transition from spots to striped structure is quite sharp in $P$.
The control parameter $P$ serves the purpose of a unique transition 
variable, having a well-defined value for the transition to occur. 
 
The most unstable mode for the smaller ($100\times100$) system  
is with $k_c=0.84$, while for the larger ($200\times200$) system  
it is with $k_c=0.45$. The larger system has smaller wave vector  
and thus the wave length, i.e., characteristic 
length of the pattern is larger. This is why the values on the 
vertical $N$-axis, are of the same order of magnitude. In order  
to compare the numbers of clusters one can normalize it by  
dividing with $N_c^d$, where $N_c = k_c L / 2 \pi$, $L$ denoting 
the linear system size (square or cube) and $d$ the spatial  
dimension. $N_c^d$ is the maximum number of spherically symmetric 
clusters in a $d$-dimensional system if the clusters were 
uniformly distributed and the effect of boundaries was neglected. 
Due to the periodicity of the chemical structure, the number of 
clusters in the actual $d$-dimensional system can be estimated to 
be $N^d = (N_c+1/2)^d$. However, an additional correction is  
required to take into account the effect of boundaries. One can  
estimate the number of additional partial clusters due to  
boundaries by estimating the length (area) of the boundary and  
the number of clusters within this domain ($d N^{d-1}$). 
As a result of this discussion we propose the normalization function 
for the number of clusters to be  
\begin{equation} 
C_{d}(N(P),N_c) = \frac{N(P)}{N_c^d}\left( 1-\frac{d}{N_c+\frac{1}{2}}\right), 
\label{eq:norm} 
\end{equation} 
where $N(P)$ is the calculated number of clusters for 
control parameter $P$. By revising the Hoshen-Kopelman algorithm 
one could have directly calculated the number of clusters by taking 
periodicity into account, in which case Eq.~(\ref{eq:norm}) reduces 
to  $C_{d}(N(P),N_c) = N(P)/N_c^d$. However, this approach was not 
implemented and the normalization was carried out by using the 
Eq.~(\ref{eq:norm}). 
 
\begin{figure} 
\includegraphics[width=.8\columnwidth, angle=270]{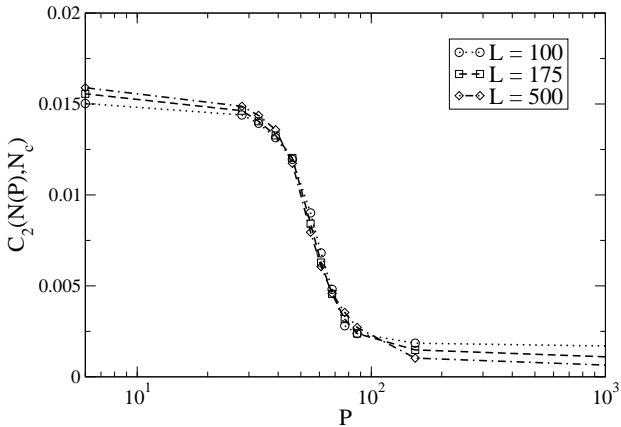} 
\caption{The normalized number of $U$ clusters as a function of $P$. 
System sizes are $L=100$ (dotted), $L=175$ (dashed) and $L=500$  
(dash-dotted).} 
\label{fig:trans1} 
\end{figure} 
 
We have also studied the effect of the system size on the mode  
transition in 2D. This is shown in Fig.~\ref{fig:trans1},  
where the normalized quantity $C_{2}(N(P),N_c)$ is plotted for  
$U$-clusters against $P$ for three different system sizes 
($L=$ 100, 175 and 500). Neglecting the number of $V$-clusters 
does not affect our conclusions, since the curves would be 
symmetrical as can be seen from Fig.~\ref{fig:result}. On the 
other hand, one can clearly see in Fig.~\ref{fig:trans1} that 
the control parameter $P$ succeeds in capturing the essential 
features of the transition. In addition, it can be seen that 
the normalization function of Eq.~(\ref{eq:norm}) scales the 
number of clusters such that it collapses onto the same curve with 
only very small deviations outside the transition. 
 
If, on the other hand, one carries out the simulations for very  
small systems, finite-size effects can be observed.  
For small system sizes the $C_{2}(N(P),N_c)$ curve becomes very  
steep in the transition region. This would suggest that in the  
limit of small systems, the transition would become almost  
discontinuous. However, the system cannot be made infinitely  
small since the (periodic) boundary conditions start to affect the  
behavior of the system. As discussed earlier the spots tend to 
nucleate from topological defects, or dislocations, of the striped 
pattern, i.e., from the points where the stripes coincide 
(Fig.~\ref{fig:sweep}). In the case of a small system even one 
dislocation can affect the morphology of the whole system and thus 
quickly transform stripes into a lattice of spots. In a larger 
system many dislocations have to appear at various sites to give 
rise to spots which in turn make the appearance of more spots favorable. 
 
So far we have discussed our simulation results in 2D systems. 
We have also studied the connectivity transition extensively in  
three dimensions. In this case stripes and spots become lamellae and 
spherical droplets, respectively, and the structure seems more 
complicated especially in the transition region. For illustrations 
of three-dimensional Turing structures we refer the reader to 
Ref.~\cite{teemu}. In 3D the transition does not occur at the same 
point with respect to $P$ as in 2D since the third dimension  
gives to the clustering process one more degree of freedom, and  
thus it is easier for the structures to connect. 
 
\begin{figure} 
\includegraphics[width=.8\columnwidth, angle=270]{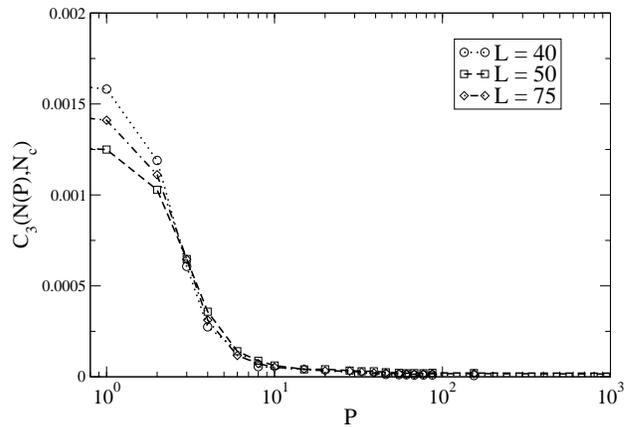} 
\caption{The normalized number of $U$ clusters as a function of 
$P$ in a three-dimensional system. System sizes are 
$L=40$ (dotted), $L=50$ (dashed), and $L=75$ (dash-dot).}  
\label{fig:trans3} 
\end{figure} 
 
This is indeed what one finds. Figure~\ref{fig:trans3} depicts  
the normalized number of clusters for three different system  
sizes ($L=$ 40, 50 and 75). One can see that the behavior of the  
system is different from two dimensions. Now, the transition  
occurs at a value of $P$ which is a decade smaller than in 2D, since 
a smaller cubic nonlinear coefficient ($r_1$) favoring lamellar  
structures is sufficient for increasing connectivity in
three-dimensional space. The significance of this large
drop of the critical $P$-value remains unanswered. 
In addition, unlike in 2D we did not observe any finite-size effects for 
the smallest possible system sizes. 
  
\section{Reciprocal space analysis} 
\label{sec:kvector} 
 
\begin{figure}[!] 
\epsfxsize=\columnwidth \epsfbox{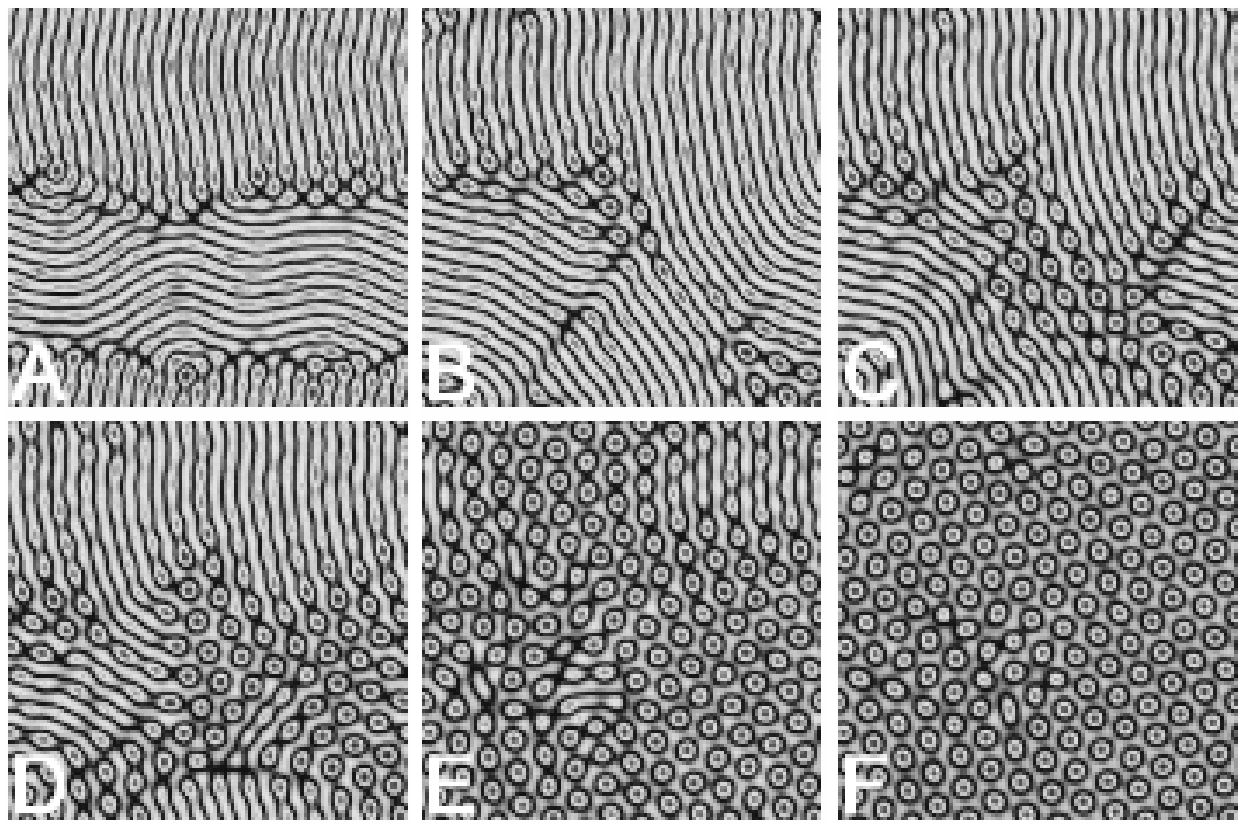}  
\epsfxsize=\columnwidth \epsfbox{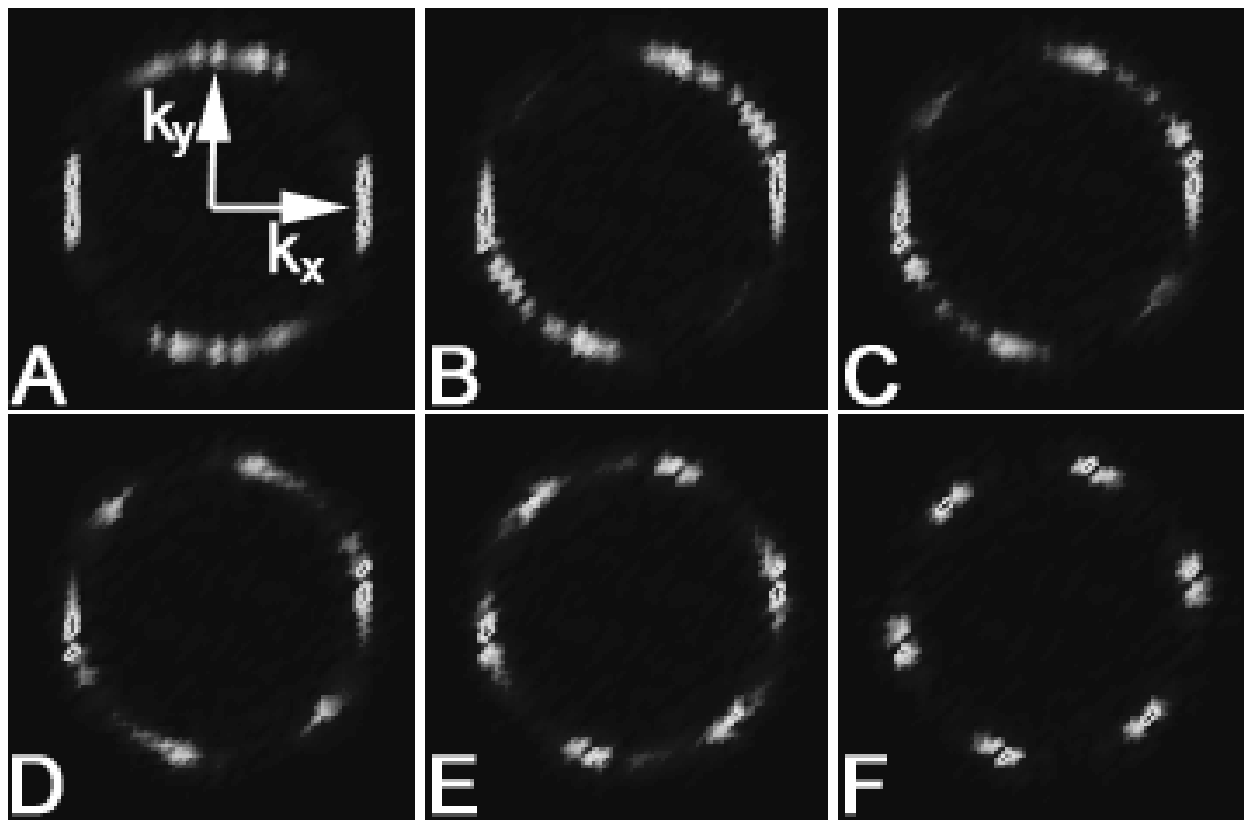}  
\caption{Chemical concentration field $\hat{\rho}(\vec{r})$
in a system of size $200 \times 200$ (top) and the corresponding 
diffraction patterns in $k_x k_y$-space (bottom)
for $P=22000$, $150$, $75$, $70$, $55$, $12$. $(k_x,k_y)=(0,0)$ 
is in the middle.} 
\label{fig:spectrum} 
\end{figure}

Next, we exploit another way to study the connectivity transition, namely  
the discrete Fourier transform of the concentration data  
in the wave vector space, $\vec{k}$, i.e.,  
\begin{equation} 
\hat{\rho}(\vec{k}) = \sum\rho(\vec{r}) e^{i\vec{k}\cdot\vec{r}}. 
\end{equation} 
This approach has been used earlier, e.g. in connection of reaction-diffusion 
systems~\cite{karttunen98a}, Turing patterns~\cite{yang}, and  
to characterize the evolution of patterns~\cite{karttunen99a}. The 
quantity $\hat{\rho}(\vec{k})$ corresponds to a diffraction pattern. 
Figure~\ref{fig:spectrum} shows a sequence of the original concentration 
fields and their diffraction patterns in a two-dimensional 
system for different control or transition field parameters, $P$.

In Fig.~\ref{fig:spectrum}A, the diffraction intensity is predominantly 
to $k_x$-direction due to more stripes in y-direction, while the smaller 
diffraction intensity around the $k_y$-axis is due to the stripes in 
x-direction. The distance of these diffraction peaks from the origin gives 
the length of the wave vector of the unstable mode, while its width in the 
perpendicular direction describes deviations of stripes from the principal 
directions due to stripe tilting and bending. In Fig.~\ref{fig:spectrum}B 
the diffraction intensity becomes more spread around the $(k_x,k_y)$ 
diagonal, as a result of appearing dislocations and nucleating spots. 
After that in Figs.~\ref{fig:spectrum}C-D the diffraction intensity 
starts splitting into separate peaks due to more spots forming, 
then developing into six separate equidistant (from the origin and 
from each other) intensity peaks as evident in Figs.~\ref{fig:spectrum}E-F. 
This hexagonal symmetry in the reciprocal space is because the system 
evolves towards regular spotty pattern with predominantly triangular 
symmetry in the real space. However, these diffraction peaks are somewhat 
spread and in fact each peak split in two, which indicates that the 
triangular symmetry is not perfect over the whole system.

If one examines the real space picture of spots in 
Fig.~\ref{fig:spectrum}F, one might think that there are fairly large 
disordered regions that deviate from the ideal triangular perfect lattice 
of spots. However, the corresponding reciprocal space picture reveals 
that this is not so, since it does not show diffuse diffraction pattern 
or a ring due to orientational disorder, but it does show a double peak 
structure as can be seen in crystalline materials due to twinning or 
low-angle grain boundaries. The existence of this kind of boundaries 
is common place in any nucleation mechanism and usually it is associated 
to the fact that twinning or low-angle grain boundaries require only 
small local displacements at low energy cost. In the present case, however, 
there is no energetics involved and in Fig.~\ref{fig:spectrum}F it is 
difficult to find clear twinning, which we found in Fig.~\ref{fig:sweep}I. 
There are other more general ways of establishing order, namely the 
appearance of {\it eutactic} local structures defined in the study of 
regular polytopes~\cite{cox}. It has been recently proposed~\cite{torres} 
that eutacticity is a very important property exhibited not only by 
crystalline and quasicrystalline lattices, but also by the geometrical 
forms of some biological systems. The presence of order in these Turing 
patterns, as revealed by their clear diffraction patterns, suggests that 
the principles governing the preference for eutactic structures also apply 
to the present case.

Here, we have done the reciprocal space analysis for 2D systems, but the 
same analysis can easily be extended to 3D systems by fixing  
the orientation of one of the spatial vectors, thus obtaining an 
in-plane diffraction graph. These diffraction patterns, however,  
are expected to look more or less the same as in 2D, which can be  
understood on the basis that almost every cross-section of a lamellae  
or spherical droplet pattern in 3D would look as striped or spotty  
pattern, respectively. 
 
\section{Discussion} 
\label{sec:concl} 
 
In this study, we have investigated the connectivity of spatial patterns  
generated by the reaction-diffusion mechanism both in 2D and 3D. This was  
done by cluster analysis for the dominating chemical. Since several  
different combinations of parameters in a Turing system can produce  
similar patterns, we derived a dimensionless control parameter 
to investigate the effects of nonlinearities. The numerical 
simulations were consistent with the predictions drawn from the 
dimensional analysis, and the system showed a transition in the  
proximity of the same $P$-value irrespective of the individual system 
parameters. With help of an analytically derived normalization function, 
the number of clusters collapsed onto the same master curve independently 
of the system size or the unstable mode. 

From biological perspective, the characteristics of the Turing  
system presented here could be important. The fact that the  
transition from striped or lamellar to spotty or spherical droplet 
structures does not depend on  the size of the system or the specific 
parameters involved, but just the combination of the nonlinearities 
($P$), makes the mechanism very general. In addition, the observation 
that irregular  structures are not as stable as regular structures 
could be used  to explain the steepness of the transition: Unfavorable 
structures  corresponding to the transition domain occupy minimal 
volume of the phase space (possible structures). 
 
The presence of eutacticity is well known in atomic crystal
structures. The fact that similar geometric principles seem to be
acting also in the formation of Turing patterns could be an important
feature in favor of their application in the study of the geometrical
forms of some simple living organisms. This could also support the idea
that Nature prefers geometrical forms generated from eutactic 
stars~\cite{torres}. This very simple analysis of Turing patterns and
their morphological transitions due to nonlinearities could open up a 
new way of investigating universal features of patterns obtained by 
very different mechanisms.
 
The approach we have taken here to study pattern formation shares  
common features with percolation and could be studied to certain  
extend as such. Percolation of morphogens to different directions 
is not as interesting a measure as the number of clusters, since 
it holds only a binary value and does not capture features of the  
morphological structure effectively. However, if one considers  
the field of biological applications, where Turing systems are  
often used, percolation behavior is interesting and has been used 
for studying some biological problems~\cite{sahimi}. 
 
If one considers the patterns in Fig.~\ref{fig:sweep}  
one can observe that in the first frame both morphogens have  
percolated in the vertical direction. On the other side of the  
transition one of the morphogens has not percolated while the  
other has done so in both directions. Should one of the chemicals  
be favorable for diffusion, these two opposite stages obtained by  
a small variation of nonlinear parameters could for example  
correspond to chemical concentrations in tissue as signaling via  
diffusion is either enabled or disabled. However, one should  
always be careful in making suggestions for biological applications  
due to complexities of nature not yet understood. 
 
\begin{acknowledgments}
We wish to thank J\'anos Kert\'esz for helpful discussions, and one of
us (R.\,A.\,B.)  wishes to thank the Laboratory of Computational
Engineering at Helsinki University of Technology for its hospitality.
This work has been supported by the Academy of Finland through its
Centre of Excellence Program (T.\,L. and K.\,K.).
\end{acknowledgments} 
 
\bibliographystyle{apsrev} 
\bibliography{turing}

\end{document}